\documentclass{LT23auth}
\usepackage{graphicx}
\usepackage{amssymb}
\usepackage{amsfonts}

\begin{document}

\begin{frontmatter}

\title{
Transmission probability through small interacting systems: \\ 
application to a series of quantum dots 
}

\author{Akira Oguri\thanksref{thank1}}

\address{Department of Material Science, 
Osaka City University, Sumiyoshi-ku, Osaka 558-8585, Japan}


\thanks[thank1]{E-mail: oguri@sci.osaka-cu.ac.jp}

\begin{abstract}
We apply the theory for 
the transmission probability $\mathcal{T}(\epsilon)$ through 
small interacting systems, 
which was formulated based on the Kubo formalism in the previous study, 
to a series of quantum dots described by the $N$-impurity Anderson model. 
Specifically, we calculate $\mathcal{T}(\epsilon)$ for $N=2$ with  
the order $U^2$ self-energy and vertex corrections  
which satisfy the current conservation, 
and examine the two different parameter regions 
at $t<\Gamma$ and $t>\Gamma$. 
Here $t$ is the inter-dot transfer and $\Gamma$ is 
the level broadening caused by
the coupling with the noninteracting leads. 
\end{abstract}

\end{frontmatter}

Recently,
we have reformulated the conductance $g$ for 
small interacting systems connected to two noninteracting leads 
based on the Kubo formula \cite{ao}. 
Carrying out the analytic continuation of the vertex corrections
 following \'Eliashberg \cite{Eliashberg}, 
we have obtained a Landauer-type expression 
applicable to the interacting electrons,  
\begin{equation}
g \  = \  {2 e^2 \over h}  
       \int_{-\infty}^{\infty} \mathrm{d}\epsilon \,
       \left( -\,{\partial f \over \partial \epsilon} \right)
        \mathcal{T}(\epsilon)
 \;,  
\label{eq:cond} 
\end{equation}
where $f(\epsilon)$ is the Fermi function.
The transmission probability $\mathcal{T}(\epsilon)$ is 
defined in terms of the vertex corrections or 
the three point correlation function \cite{ao},
and depends on temperature $T$ in the interacting systems. 
Our derivation uses neither the precise form 
of the Hamiltonian nor the perturbation expansion,
and thus the formulation can be applied to various systems 
such as quantum dots and atomic wires of nanometer size.

In this report 
we will apply the theory to the series of $N$ Anderson impurities, 
which can be regarded as a model for a network of quantum dots,
and examine the parameter region that was not studied  
in the previous paper \cite{ao}.
The system we consider consists of three regions;
a small interacting region at the center ($C$),
 and two noninteracting leads at left ($L$) and 
right ($R$). 
The total Hamiltonian is 
$\mathcal{H}_{\mathrm{tot}}  
=   \mathcal{H}^0 + \mathcal{H}_{C}^{\mathrm{int}}
$,
with 
$
\mathcal{H}^0  
=  \mathcal{H}_{\mathrm{L}} + \mathcal{H}_{R}  
+ \mathcal{H}_{C}^0 + 
\mathcal{H}_{\mathrm{mix}} 
$.
Here, $\mathcal{H}_{L}$ and $\mathcal{H}_{R}$ are 
the Hamiltonian for the noninteracting leads.  
The central region and the connection to the leads
are described by
\begin{eqnarray}
\mathcal{H}_{C}^0 \, &=& \, - t\,
\sum_{j=1}^{N-1} \left(\,
 c^{\dagger}_{j+1 \sigma} c^{\phantom{\dagger}}_{j \sigma}  
+ \mathrm{H.c.} \,\right)    
\;,
\\
\mathcal{H}_{C}^{\mathrm{int}} \, &=& \, U \sum_{j=1}^N 
\Bigl[\, 
n_{j \uparrow}\, n_{j \downarrow}-(n_{j \uparrow}
+ n_{j \downarrow} )/2 \, \Bigr] \;,
\\
\mathcal{H}_\mathrm{mix}  \, &=& \, 
   -v \,\sum_{\sigma}\left(\, 
   c^{\dagger}_{1 \sigma} c^{\phantom{\dagger}}_{0 \sigma}
  +   
   c^{\dagger}_{N+1 \sigma} c^{\phantom{\dagger}}_{N \sigma}
          +  \mbox{H.c.} \,\right)  \;,
\end{eqnarray}
in the standard notation.
Here the label $0$ ($N+1$) assigned to the site 
at the interface of the left (right) lead.
The contributions of $\mathcal{H}_\mathrm{mix}$ are described through  
the parameter $\Gamma = \pi \rho\, v^2$, where $\rho$ is 
the density of states of the isolated lead.

Specifically, in this report 
we consider the electron-hole symmetric case 
and focus on the system of $N=2$.
The conductance $g$ for the two-impurity Anderson model 
has been studied extensively. 
Particularly, the competition between the Kondo effect 
and the inter-dot magnetic exchange coupling 
occurring in this system has been investigated 
in detail based on the precise calculations by the 
numerical renormalization group method \cite{Izumida}. 
Nevertheless, the transmission probability $\mathcal{T}(\epsilon)$ has 
not been studied systematically, and as seen below 
it contains important information about the excitation spectrum.

\begin{figure}[t]
\begin{center}
\leavevmode
\includegraphics[width=1.025\linewidth, 
clip, trim = 0.5cm 11.75cm 0cm 11.25cm]{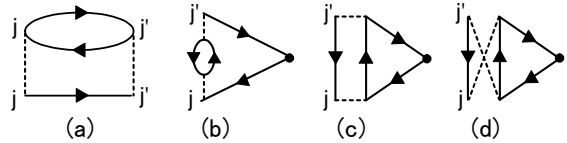}
\caption{The order $U^2$ corrections of (a) the self-energy 
and (b)--(d) the current vertex, where $1\leq j,j' \leq N$.} 
\label{fig:diagram}
\end{center}
\end{figure}

\begin{figure}[t]
\begin{center}
\leavevmode
\includegraphics[width=1.07\linewidth, 
clip, trim = 0.3cm 0.8cm 0cm 0cm]{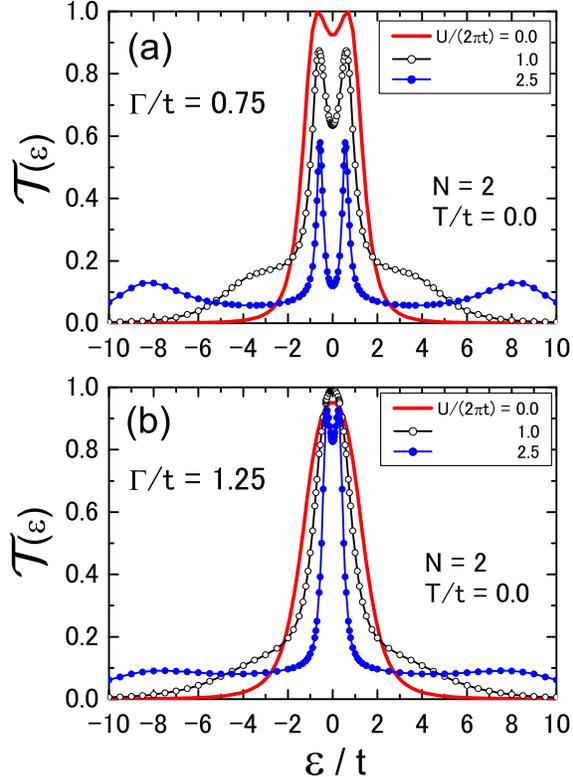}
\caption{
Transmission probability for $N=2$ at $T=0$ 
for three values of $U/(2\pi t)$;
(---) $0.0$, (--$\circ$--) $1.0$, and (--$\bullet$--) $2.5$.
The value of the ratio $\Gamma/t$ is taken 
to be (a) $0.75$ and (b) $1.25$. 
}
\label{fig:trans}
\end{center}
\end{figure}

To calculate $\mathcal{T}(\epsilon)$ from the expression   
which is written in terms of the Green's functions 
and vertex corrections \cite{ao}, 
we take into account the order $U^2$ self-energy and vertex part 
illustrated in Fig.\ \ref{fig:diagram}, taking $\mathcal{H}^0$ to be 
the unperturbed Hamiltonian.  
We note that the current conservation is satisfied 
in the calculations. 
In Fig.\ \ref{fig:trans} the transmission probability at $T=0$ is  
plotted for $N=2$, where $U/(2\pi t)$ is chosen to be 
(---) $0.0$, (--$\circ$--) $1.0$, and (--$\bullet$--) $2.5$. 
The ratio of the mixing to the inter-dot transfer, 
$\Gamma/t$, is taken to be (a) $0.75$ and (b) $1.25$.

The low-energy behavior of $\mathcal{T}(\epsilon)$ is 
quit different depending on the  value of $\Gamma/t$.
As seen in Fig.\ \ref{fig:trans} (a), 
for $\Gamma <t$ there are two resonant peaks at $\epsilon \simeq \pm t$,
which for $U=0$ correspond to the bonding and anti-bonding states 
of the two dots.
As $U$ increases, 
the height of these two low-energy peaks decreases, 
but the width of the peaks becomes sharp and the valley 
at $\epsilon =0$ becomes deep. 
Two additional broad peaks at high energy $\epsilon \simeq \pm U/2$ 
appearing for $U/(2 \pi t) \gtrsim 1.0$  correspond 
to the upper and lower Hubbard levels of the atomic character. 
 Fig.\ \ref{fig:trans} (b) shows the results 
in the other parameter region $\Gamma >t$. 
For $U/(2\pi t) \lesssim 1.0$, the transmission probability 
$\mathcal{T}(\epsilon)$ has only a single peak at low energies. 
The hight of this low-energy peak increases with $U$ until 
it reaches the unitary limit value. 
Then for $U/(2 \pi t) \gtrsim 1.0$ the hight of the peak  
decreases with increasing $U$, 
and eventually the single peak splits into two 
as seen in the results for $U/(2 \pi t) =2.5$.

The conductance can be obtained from eq.\ (\ref{eq:cond}), 
and at $T=0$ it is determined by the value 
of $\mathcal{T}(\epsilon)$ at $\epsilon=0$. 
Thus for $\Gamma <t$ the conductance decreases 
monotonically with increasing $U$, 
while for $\Gamma >t$  it shows a maximum 
at the value of $U$ which corresponds to the unitary limit. 
At low temperatures, 
the $T$ dependence of the conductance 
is scaled by the energy determined by the width of the valley 
when $\mathcal{T}(\epsilon)$ has 
the two-peak structure at low energies. 
In the other case, for the single-peak structure,  
the energy scale is determined by the width of the central peak. 
We note that the temperature dependence of 
$\mathcal{T}(\epsilon)$ arises through  
that of the self-energy and vertex corrections, 
and at high temperatures the peak structure at low energies 
is smeared by the thermal fluctuations 
as demonstrated for $\Gamma/t=0.75$ in the previous paper \cite{ao}.

The transmission probability $\mathcal{T}(\epsilon)$ we have 
formulated can also be written in terms of the three-point 
correlation function \cite{ao},  
and the Lehmann representation of it can be 
used for the nonperturbative approaches  
such as the numerical renormalization group 
and the quantum Monte Carlo methods.

\begin{ack}

\vspace{-0.25cm}

We would like to thank H. Ishii for valuable discussions. 
Numerical computation was partly carried out 
 at Yukawa Institute Computer Facility.
This work is supported by the Grant-in-Aid 
for Scientific Research from the Ministry of Education, 
Science and Culture, Japan.
\end{ack}

\end{document}